# Evaluating the role of risk networks on risk identification, classification and emergence

Christos Ellinas[1,2,*], Neil Allan[1,2], Caroline Coombe[3]

[1] University of Bristol, Bristol, UK

[2] Systemic Consult Ltd, Bradford-on-Avon, UK

[3] ORIC International, UK

[*]Corresponding author

E-mail: ce12183@bristol.ac.uk

## Abstract

Modern society heavily relies on strongly connected, socio-technical systems. As a result, distinct risks threatening the operation of individual systems can no longer be treated in isolation. Consequently, risk experts are actively seeking for ways to relax the risk independence assumption that undermines typical risk management models. Prominent work has advocated the use of risk networks as a way forward. Yet, the inevitable biases introduced during the generation of these survey-based risk networks limit our ability to examine their topology, and in turn challenge the utility of the very notion of a risk network. To alleviate these concerns, we proposed an alternative methodology for generating weighted risk networks. We subsequently applied this methodology to an empirical dataset of financial data. This paper reports our findings on the study of the topology of the resulting risk network. We observed a modular topology, and reasoned on its use as a robust risk classification framework. Using these modules, we highlight a tendency of specialization during the risk identification process, with some firms being solely focused on a subset of the available risk classes. Finally, we considered the independent and systemic impact of some risks and attributed possible mismatches to their emerging nature.

## Key messages

- The authors introduce a novel methodology to generate risk networks, which relaxes some assumptions of past related work, including prominent work by the World Economic Forum.
- The risk network can serve as a robust risk classification framework, free from externally-imposed artificial constructs (e.g. risk classification based on regulatory requirements).

- By applying it to a (re)insurance dataset, the authors use the resulting risk network to evaluate the 'horizon scanning' capacity of every included firm.
- The risk network uncovers a general mismatch between independent and systemic impact of risks, with authors arguing that this mismatch can be used to classify certain risks as 'emerging'.

## Introduction

Enhanced understanding on the nature of risk is the epitome of modern science (Bernstein & Bernstein Peter, 1996; Buchanan & O Connell, 2006), with its successful management yielding significant benefits across a wide range of societal facets (Ganin et al., 2016; Helbing, 2013; Vespignani, 2012). In this context, risk is traditionally defined as the "effect of uncertainty on objectives"; it is generally quantified as the probability of an event materializing, times its expected impact (ISO, 2009). The objective of risk management is thus, to mitigate against events that can lead to an undesirable outcome (Pritchard, 2014).

Underlying this objective is the assumption that each such adverse event is independent e.g. interdependence bears no effect when quantifying risk (ISO, 2009). Yet, the operation of modern society largely depends on precisely this interdependence (World Economic Forum, 2017), as it supports the global exchange of "people, goods, money, information, and ideas" (Helbing, 2013). Incorporating the effect of interdependence into the risk management process has attracted much recent interest (Battiston, Caldarelli, D'Errico, & Gurciullo, 2016; Battiston, Puliga, Kaushik, Tasca, & Caldarelli, 2012; DasGupta & Kaligounder, 2014; Helbing, 2013; Roukny, Bersini, Pirotte, Caldarelli, & Battiston, 2013; Szymanski, Lin, Asztalos, & Sreenivasan, 2015) partly due to the 2007 – 2008 global financial crisis and the way in which traditional risk models, also grounded on the assumption of risk independence, failed to foresee it (Battiston, Caldarelli, et al., 2016; Battiston, Farmer, et al., 2016; Besley & Hennessy, 2009; Schweitzer, Fagiolo, Sornette, Vega-Redondo, & White, 2009).

One way of exploring the effect of risk interdependence is by considering how risks interact (Helbing, 2013; Szymanski et al., 2015). A prominent example of this approach can be found in the annual 'Global Risk Report', generated by the World Economic Forum (WEF). Currently in its 12$^{th}$ edition, this report explicitly explores the effect of risk interdependence by considering *risk networks*. Within each network, a risk (node) is connected, via weighted links, to a number of other risks. In this particular example, links are established through a survey of roughly 750 experts – from government, academics and industry – with participants being asked the following: "Global risks are not isolated and it is important to assess their interconnections. In your view, which are the most strongly connected global risks? Please select three to six pairs of global risks?" (see Appendix B in (World Economic Forum, 2017)).

This question focuses on describing the local structure of the risk network, and is a variant of the so-called 'name generator' – a tool often deployed by surveys which focus on constructing the overall structure of (mostly social) networks using ego networks (Bidart & Charbonneau, 2011; Merluzzi & Burt, 2013). Despite the wide deployment of these 'name generators' (Merluzzi & Burt, 2013), the resulting data must be approached with caution due to their inevitable exposure to multiple sources of contamination (Bearman & Parigi, 2004; Bidart & Charbonneau, 2011; M. Newman, Barabasi, & Watts, 2011). In the case of the WEF report, the derived risk network must be regarded with skepticism, for at least two reasons. Firstly, participants are explicitly given an upper and lower bound on the number of connections that they can utilize, inevitably biasing the overall connectivity of the risk network. Secondly, the nature of the link implied through the questionnaire is ambiguous, as a link between two risks may suggest: (a) a causal link (i.e. Risk A *causes* Risk B and hence, they are connected) or, (b) a similarity link (i.e. Risk A is *similar* to Risk B, and hence, they are connected). This accumulating ambiguity can undermine the consequent analysis of the resulting risk network. For example, consider the most connected risk. If (a) is the case, then this risk is expected to play a key role in terms of triggering large-scale cascades i.e. be of high systemic importance (Albert & Barabási, 2002). Yet if (b) is the case, such heightened connectivity merely suggests that its neighboring risks are somewhat similar. The impact of this ambiguity becomes even harder to evaluate once sophisticated analysis is applied on top of such networks. For example, consider the recent work of Szymanski et al. (2015), who have used the WEF risk network to analyze its failure dynamics. Despite the theoretical rigor of the analysis itself, its inevitable dependence on the network's topology questions the eventual outcome of the analysis, since the ambiguity contained within the network itself is neither evaluated nor accounted for.

Working towards capturing risk interdependence in a more robust way, we developed a methodology to generate weighted risk networks based on risk similarity, where risks are connected based on the similarity of their characteristics. By applying this methodology to an empirical dataset of 143 risks – each described using 24 unique tags – this paper discusses the role of risk interdependence in terms of three core components of the risk management process: (a) risk classification which is independent of externally-imposed labels; (b) evaluation of the 'horizon scanning' capacity of a given firm; (c) identification of emerging risks – based on the influence of interconnectivity on their independent impact – and how they underlie firm interactions.

## Results

In what follows, we analyze the topology of the risk network, particularly focusing on its modular composition (see Supporting information (SI) for detailed visualizations). We then evaluate the capacity of each firm to identify risks uniformly across all observed modules – an ability to do so corresponds to an enhanced 'horizon scanning' capacity. Finally, we use a simple epidemic model (Gutfraind, 2010;

Pastor-Satorras, Castellano, Van Mieghem, & Vespignani, 2015; Watts, 2002) to evaluate the systemic importance of each risk, in terms of its ability to trigger subsequent risks. By doing so, we compare the reported independent impact of each risk with its evident systemic one, attributing possible differences to their 'emerging' nature. The consequent interaction between firms is briefly evaluated, in the form of liability networks.

**Emergence of risk modules**

In the context of the risk network, our analysis identifies five distinct modules, composed of 47 (Module 1), 35 (Module 2), 25 (Module 3), 21 (Module 4) and 16 (Module 5) risks respectively – see Figure 1. A module is defined as a group of nodes densely connected between each other yet loosely connected with nodes that belong to different modules (Danon, Diaz-Guilera, Duch, & Arenas, 2005; Fortunato, 2010). In the context of the risk network, every module can be regarded as a distinct 'risk class', where its formation solely depends on the underlying characteristics of each risk (see Method). This bottom-up approach contrasts typical risk classification schemes, where a top-down approach is generally adopted, building on externally-imposed labels i.e. based on a particular organizational function, e.g. 'Strategic Risk' (Kaplan & Mikes, 2012), or a regulatory requirement, e.g. 'Capital Ratio' from the Basel III regulatory framework (Basel Committee, 2010)).

Increased levels of connectivity correspond to increased levels of risk similarity, both in terms of intra-connectivity (*within* a module) and inter-connectivity (*between* modules). Consequently, if a given set of conditions triggers a particular risk, the same condition(s) will also affect (and potentially trigger) its neighboring nodes, depending on how similar they are in terms of their underlying characteristics (Allan, Cantle, Godfrey, & Yin, 2013). With risk similarity in mind, consider the case of Module 2 – heightened intra-connectivity indicates that the risks contained within it are increasingly similar; conversely Module 3 is defined by relatively low levels of intra-connectivity. Shifting focus to the inter-connectivity aspect, heightened inter-connectivity identifies related risk classes – the strong link between Module 2, composed of regulatory risks, and Module 5, composed of political risks, serves as an intuitive example.

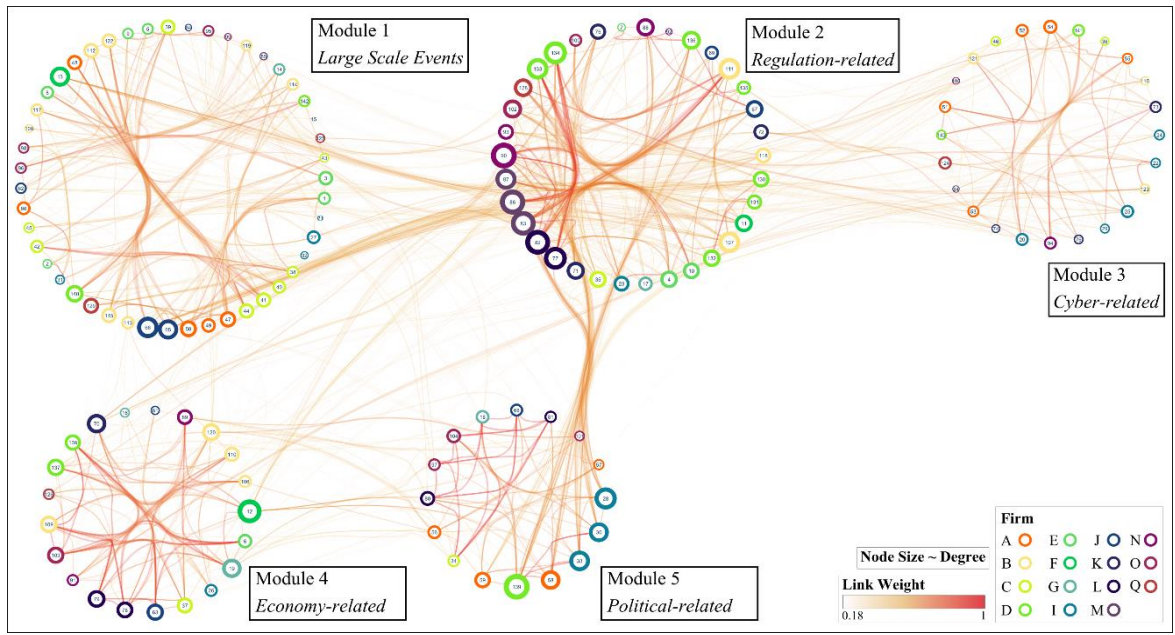

**Figure 1**: Risk Network, where a weighted link between two nodes (risks) reflects their similarity, in terms of their risk characteristics. Node color corresponds to the firm who reported it; node size corresponds to its degree and link color corresponds to link weight. Additional information can be found in (ORIC International, 2017).

With respect to the actual composition of each module, it is of particular interest to identify risks that contradict the overall theme of each module. For example, Module 3 is principally composed of cyber-related risks, evident by the word-decomposition of the risk labels found within the module (see Section 2, SI). Amongst these cyber-related risks, the risk 'Global Population Changes' is also present – a rather counter-intuitive inclusion at first, yet one that is deeply embedded within the technological realm of Module 2. One can easily reason that health is highly dependent on the rate of technological advancement, which in turn affects the population size. Yet in the case of traditional risk classification, 'Global Population Change' would have been grouped under a distinctly different label (e.g. 'Insurance and Demographic Risk' (Kelliher, Wilmot, Vij, & Klumpes, 2013)) compared to the rest of the risks contained in Module 2. More generally, these risk modules can uncover risks which are seemingly distinct in terms of their attributed label – such as the case with 'Global Population Changes' – yet are increasingly similar in terms of their underlying characteristics, which in turn suggests some sort of similarity in the way for mitigating against them.

**Evaluation of 'Horizon Scanning' capacity**

The risk management process can be summarized as a process aimed to "Identify to Analyze to Evaluate to Treat" a particular risk (ISO, 2009). With 'horizon scanning' being the first step in this process, a firm capable of identifying risks across all 'risk classes' limits its exposure to unidentified risks. By considering the basis on which the network is developed, this insight becomes intuitive: when a firm identifies, and eventually treats, a risk of a given class the firm inevitably becomes somewhat shielded from the impact of similar risks i.e. risks that belong to the same class (World Economic Forum, 2017). Conversely, the tendency of a firm to identify risks from particular 'risk classes' biases its 'horizon scanning' function, and in turn, increases its overall risk exposure, especially if entire 'risk classes' remain uncovered.

Table 1 details the 'horizon scanning' capacity of each firm, as reflected by the number of risks identified in each of the five 'risk classes' (reported in the form of a percentage). An example of the aforementioned bias towards missing particular 'risk classes' is Firm A, with its 'horizon scanning' deployment specializing in risks that belong to Module 1, 3 and 5 (Figure 2; blue). As a result, Firm A is unaware of the risks that belong to Module 2 and 4. Conversely, Firm O is able to identify at least one risk across all five modules (Figure 2; red), and hence is well-equipped to tackle risks which have remained unidentified yet are contained within these five modules.

More generally, the majority of firms appear to specialize in the identification of risks that belong to particular risk classes. In other words, firms tend to tailor their 'horizon scanning' function towards the identification of risks of a particular nature (i.e. risks that belong to the same module). Network-based techniques can highlight these instances and help mitigate them by broadening the focus of the corresponding 'horizon scanning' function.

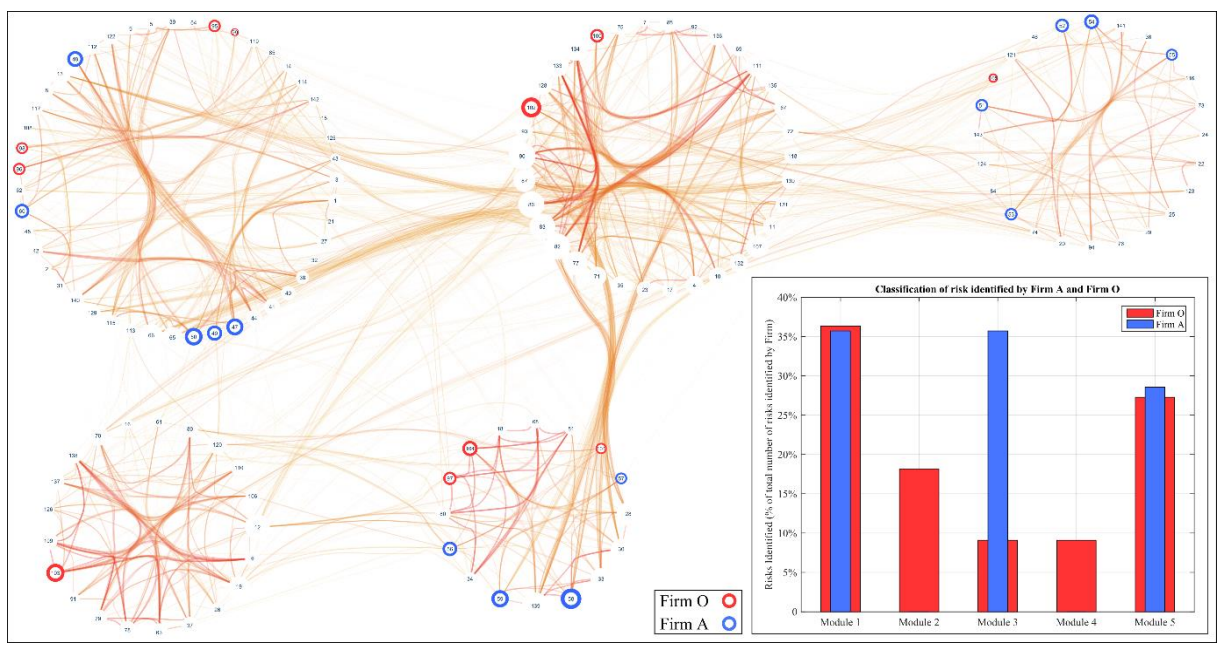

**Figure 2**: 'Horizon scanning' capacity of Firm O (red) and Firm A (blue), in terms of identifying risk across the network. Firm O successfully identifies at least one risk from each module, while Firm A is limited to Module 1, 3 and 5. Subplot includes the corresponding histogram. Additional information can be found in (ORIC International, 2017).

**Table 1**: Percentage of risk(s) identified by each firm across the five modules. Each percentage corresponds to the number of risks identified over the total number of risks reported by the corresponding firm. Results are rounded to 1 decimal place. Note that every Firm ID corresponds to a firm contained within the dataset, anonymized for confidentiality purposes.

| Firm | Module | | | | |
|---|---|---|---|---|---|
| | 1 | 2 | 3 | 4 | 5 |
| A | 35.7% | 0.0% | 35.7% | 0.0% | 28.6% |
| B | 44.4% | 16.7% | 16.7% | 22.2% | 0.0% |
| C | 61.5% | 7.7% | 15.4% | 7.7% | 7.7% |
| D | 14.3% | 50.0% | 14.3% | 14.3% | 7.1% |
| E | 60.0% | 30.0% | 0.0% | 10.0% | 0.0% |
| F | 33.3% | 33.3% | 0.0% | 33.3% | 0.0% |
| G | 33.3% | 16.7% | 0.0% | 33.3% | 16.7% |
| I | 28.6% | 7.1% | 35.7% | 7.1% | 21.4% |
| J | 44.4% | 22.2% | 0.0% | 22.2% | 11.1% |
| K | 0.0% | 42.9% | 42.9% | 14.3% | 0.0% |
| L | 0.0% | 33.3% | 0.0% | 33.3% | 33.3% |
| M | 20.0% | 60.0% | 20.0% | 0.0% | 0.0% |
| N | 0.0% | 57.1% | 14.3% | 28.6% | 0.0% |
| O | 36.4% | 18.2% | 9.1% | 9.1% | 27.3% |
| Q | 33.3% | 33.3% | 16.7% | 16.7% | 0% |

**Identifying emerging risks and who they affect**

An emerging risk can be defined as "a material, previously unconsidered risk or changing risk factor that has the potential to significantly alter the firms' risk profile" (ORIC International, 2017). These risks are "developing or already known risks which are subject to uncertainty […] and are therefore difficult to quantify using traditional risk assessment techniques" (International Actuarial Association, 2008). In this context, we translate this uncertainty as the way in which interconnectivity affects the systemic impact of a risk, in relation to its independent impact. In other words, an emerging risk is one in which its position in the network alters its independent impact, either in a positive *or* a negative impact.

The independent impact of every risk considered herein is reported in a qualitative manner (i.e. 'High', 'Medium' or 'Low') by its respective firm, as set by the industry standard ISO 31 000 (ISO, 2009). The systemic impact is evaluated using a simple threshold model (Gutfraind, 2010; Pastor-Satorras et al., 2015; Watts, 2002), which essentially models a cascade in which a risk materializes, and subsequently triggers related risks in a probabilistic manner, depending on the risk similarity of a risk pair (see

Method). The final number of risks consequently affected by the initially affected risk corresponds to its systemic impact. As such, a risk whose materialization in turn triggers a large number of subsequent risks is assigned a high systemic impact, and vice-versa.

Overall, a general mismatch exists between the independent and systemic impact across the most influential risks, which indicates that the assumption of risk independence obscures the emerging nature of risks (Figure 3). This misalignment is consistent across most firms, highlighting an overall tendency to underestimate the increased systemic impact of particular risks. Consider Risk IDX 118 ('European data protection rules'), which has been assessed to have a 'Low' independent impact, yet it is of 'High' systemic impact (triggers an average of 32.9 subsequent risks; ranks 4th out of the 143 risks) – see Table 2. In other words, the assumption of risk independence conceals the systemic nature of these risks, and in turn shrouds its emerging nature.

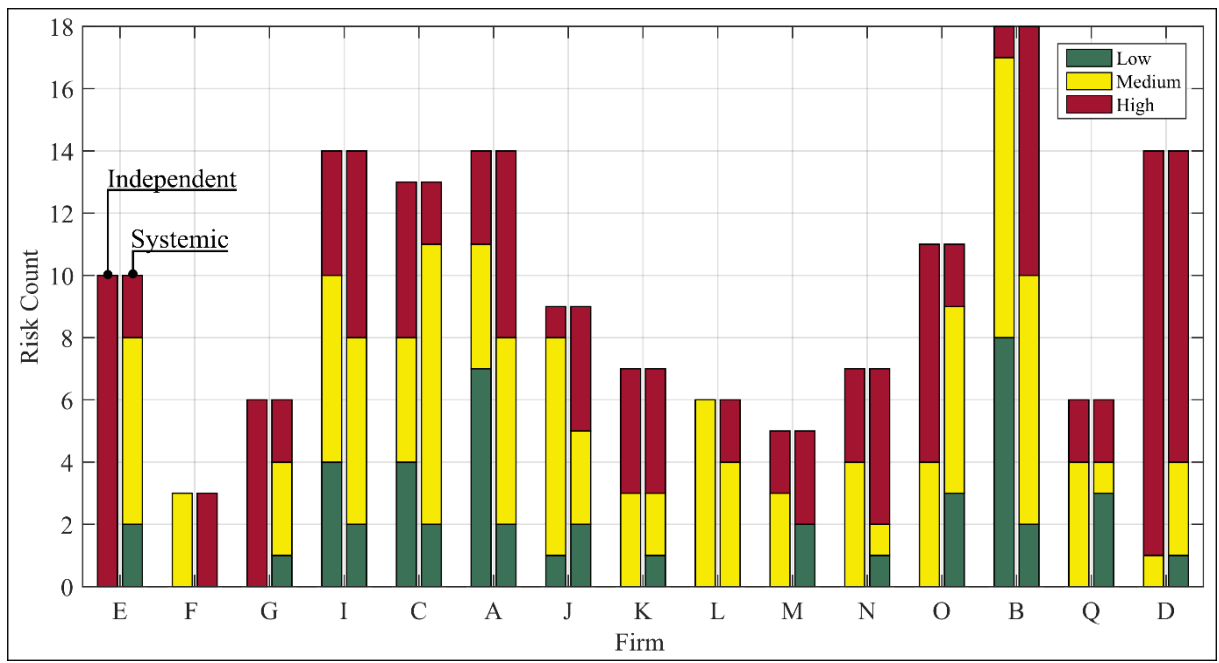

**Figure 3:** Breakdown of independent (1st column) and systemic impact (2nd column) for every risk, grouped by the firm who has reported it. Bar color corresponds to the size of impact. Additional information can be found in (ORIC International, 2017).

**Table 2**: Top five risks in terms of their systemic impact. Independent impact is also included, along with the firm which has reported each risk. Note that every Firm ID corresponds to a firm contained within the dataset, anonymized for confidentiality purposes.

| Trigger Risk IDX | Risk Title | Firm who identified Risk | Systemic Impact (Rank) | Independent Impact |
| --- | --- | --- | --- | --- |
| 65 | 'Political intervention (tax, caps, levies, rating factors, data etc.)' | 'J' | High (1$^{st}$) | Medium |
| 12 | 'Brexit and Scottish independence' | 'F' | High (2$^{nd}$) | Medium |
| 107 | 'Legal action driving changing claims patterns' | 'B' | High (3$^{rd}$) | Medium |
| 118 | 'European data protection rules' | 'B' | High (4$^{th}$) | Low |
| 13 | 'Mandate extension to commercial properties' | 'F' | High (5$^{th}$) | Medium |

The consequent interaction between firms, as it emerges through the systemic nature of each risk, can be examined by considering the liability network. In this case, each node corresponds to a firm, and a link between firm $i$ and $j$ reflects the ability of at least one risk reported by firm $i$ to interact with at least one risk reported by firm $j$. In addition, link weight corresponds to the number of times all risks reported by firm $i$ interact with risks reported by firm $j$ (Figure 4). This weight is normalized over the total number of risks reported by firm $i$ in order to account for the variability in the number of risks reported by each firm. Note that despite the symmetry in link directionality, this normalization scheme allows for a link between firm $i$ and $j$ to be of different weight compared to the link between firm $i$ and $j$; thus we consider the liability network to be directed.

The liability network can be used to identify firms that are heavily exposed to the systemic impact of particular risks, and highlight possible collaborations. For example, Firm D is the one most affected, evident by the largest weighted in-degree (proportional to node size; Figure 4, Left Panel). In addition, the largest contribution comes from risks that have been reported by Firm F. In other words, risks which have been reported by Firm F are very similar to the ones reported by Firm D, and in turn are increasingly likely to affect the former (Firm F). Similarly, risks reported by Firm F have a high systemic impact (Figure 4; Right Panel); making Firm F a key collaborator from which information sharing can benefit affected firms, such as Firm D. Therefore, one can envision a collaboration between Firm F and D in an attempt to prevent risk more efficiently.

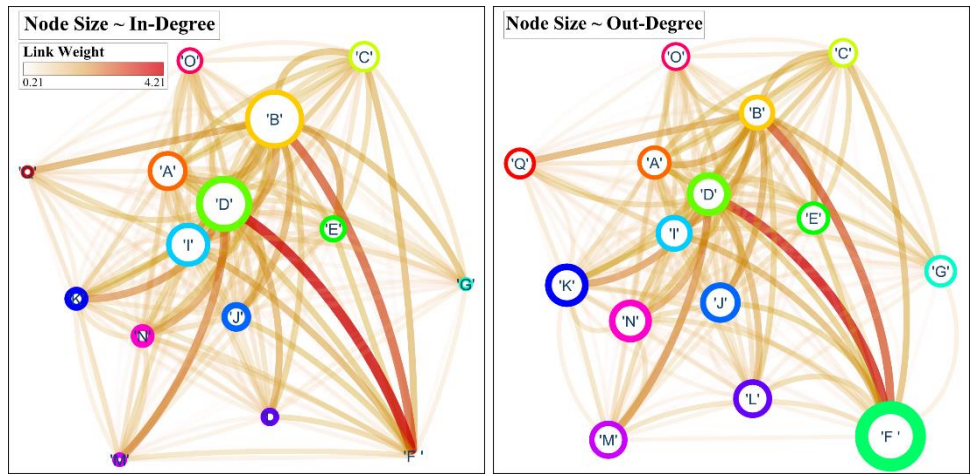

**Figure 4:** Liability network, where each node corresponds to a firm, and a link between firm $i$ and $j$ reflects the capacity of at least one risk by firm $i$ to interact with a risk reported by firm $j$. The weight of a link from node $i$ to node $j$ corresponds to the total number of times a risk identified by firm $i$ has affected a risk identified by firm. Node size is proportional to the in-degree (Left Panel) and out-degree (Right Panel) of each node.

## Discussion

In this paper, we have presented an evaluation of how risk interdependence affects the risk management process. In contrast to previous studies, which focus on survey-based risk networks, we have introduced an empirical-based, quantitative risk network. In this respect, we have focused on: (a) the emergence of network modules, (b) the 'horizon scanning' capacity of individual firms, and (c) emergent risks and how they reflect firm interactions.

Modules within the risk network provide an intuitive way for classifying risk. Typically, risk classification takes place within the boundary of individual firms through the imposition of, what is hoped to be, meaningful labels. Each such label relates a particular aspect of a firm to its economic value e.g. 'Market Risk' relates market movement to fluctuations in the value of existing assets, which in turn impacts a firm's liabilities and income. Yet such classification is driven by externally-imposed labels which can fuel ambiguity, resulting in similar risks being grouped differently. A recent report by the Institute and Faculty of Actuaries highlights this inconsistency by the means of an example, where "one organization may class failure of a project as Operation Risk, while another class it as Strategy Risk" (Kelliher et al., 2013). Transitioning from high-level risks – such as the ones considered herein – to low-level risks fuels the frequency of such inconsistencies further, as the number of possible labels that can be attributed to any one risk explodes (Kelliher et al., 2013).

In contrast, the methodology proposed herein provides an intuitive way for classifying risk. By looking beyond a risk's label, the explicit focus on a risk's underlying characteristics ensures that the classification process is not obscured by externally imposed labels. Rather, the focus is on risk similarity, ensuring that risks belonging to the same module are in fact, alike. As a result, these modules can include risks that are similar in principle, yet described by seemingly unrelated labels with respect to the rest of the module. Consequently, resources spent in managing risks that appear to be different, yet are fundamentally similar – i.e. they belong to the same risk module – can be saved, effectively streamlining the risk management process.

By exploiting the emergence of these modules, a firm can navigate towards enhanced 'horizon scanning' capabilities, by identifying a diverse set of risks i.e. across all identified network modules. Considering the similarity-based construction of the risk network, the ability to identify risks from each module suggests that even though a firm may have missed some risks, its overall preparedness is high as the remaining risks within that module are similar in nature. Overall, our work shows that the majority of firms specialize in the identification of risks that are of similar nature (i.e. risks that belong to the same 'risk class'). Whilst such specialization is understandable, it can also increase risk exposure due to unidentified risks creeping in. Introducing network-based techniques into the overall risk management process can help contain this effect, improving the overall effectiveness of the risk management process.

Finally, we consider the effect of interconnectivity in terms of a possible mismatch between the independent and systemic impact of any given risk – we refer to risks that exhibit this mismatch as emerging risks. Focusing on risks where interconnectivity has a worsening effect, we are able to identify risks with small independent impact, yet capable of larger systemic impact. Such insight can be used to minimize biases introduced by traditional tools (e.g. risk registers (ISO, 2009)), where attention is skewed towards risks with a high independent impact. In doing so, the likelihood of omitting risks with low independent impact, yet potentially high systemic impact can be minimized. In addition, by translating the systemic impact of each risk into a liability network, we can identify beneficial collaborations between firms, where the neighbor of a firm can hold valuable information with respect to the risks that impact it (e.g. Firm D and F; Figure 4). In principle, one could envision such information to be used in order to promote mutually beneficial collaborations that can increase risk mitigation efficiency.

With that in mind, it is worth highlighting that firms are complex, multi-faceted systems operating across a wide range of environments (e.g. regulatory, commercial etc.). Therefore, the utility of the liability network in identifying joint exposures that emerge from this rich variety of dependencies depends on *a priory* information. Consider a simple example where contractual dependencies have been analyzed, and the risk of heavy reliance to a particular partner identified. On the one hand, if this risk is appropriately recorded, then it's contribution to the liability network will be present; on the other hand, if the risk has been omitted then the liability network will be inevitably incomplete, and hence its utility diminished.

In conclusion, the use of quantitative risk networks can significantly contribute to the spurring discussion on the interdependent nature of risk, and its effect. The ability to map risk interdependence in the form of network modules provides for a natural way to classify risk, which in turn can provide an intuitive way to reduce the number of risks that can be managed from several hundred to a handful, focusing the risk management effort. With that in mind, strategies can be formulated to prevent the occurrence of multiple risks that belong to the same class, and therefore increase effectiveness and efficiency of the overall risk management process. In addition, the capacity to evaluate possible limitations in the 'horizon scanning' capacity of a firm can provide valuable insight on possible exposures, whilst the capacity to identify emerging risks contributes to the reduction of a firm's exposure to large-scale, systemic failures.

## Methods

### Data

The risk dataset has been obtained from ORIC International – an operational risk consortium for the (re)insurance and asset management sector (www.oricinternational.com). The dataset contains 143 unique risks, as reported from 15 firms active in the (re)insurance sector.

Every risk $i$ is characterised by a row vector $c^i = c_1^i, c_2^i, c_3^i, \ldots, c_{24}^i$ where each entry is binary and reports whether a particular theme tag is present. The set of risk characteristics considered is provided in Table 3; the raw data is available in the SI.

**Table 3**: Set of characteristics used to describe the 143 risks.

| Risk Characteristics | | | | |
|---|---|---|---|---|
| 1. 'Natural Disasters' | 2. 'Pandemic/ Health' | 3. 'Underwriting experience' | 4. 'Statutory / Regulatory changes' | 5. 'DR/BCP' Business contingency planning |
| 6. 'Investments' | 7. 'Capital modelling' | 8. 'Political instability' | 9. 'Climate' | 10. 'Outsourcing' |
| 11. 'Cyber' | 12. 'Technology/ Data' | 13. Competition/ Distribution channels' | 14. 'Consumer behavior' | 15. 'Terrorism/ War' |
| 16. 'Credit/ Market shocks' | 17. 'Operational disruption' | 18. 'War/Terrorism' | 19. 'Claims' | 20. 'Pricing' |

| 21.'Customer service' | 22.' Crime' | 23.' Reputation ' | 24.' Data' |

---

**Risk Network Generation**

In general, a network $G(N, E)$ is composed of a set of nodes $N$, $N \equiv \{n_1, n_2, ..., n_N\}$ and edges $E, E \equiv \{e_1, e_2, ..., n_E\}$. The structure of the network is stored in an $N \times N$ matrix, called the adjacency matrix, **A**. A non-zero $\mathbf{A}(i,j)$ entry corresponds to a link between node $i$ and $j$ with a weight equal to the magnitude of the entry.

To generate a risk network, we first construct a similarity matrix **S**, where $\mathbf{S}(i,j)$ records the similarity between risk $i$ and $j$ (see Section 3, SI). This similarity is quantified using the Cosine Distance between the two corresponding characteristics vectors $c^i$ and $c^j$, defined as:

$$d_{i,j} = \frac{c^i \cdot c^j}{\sqrt{(c^i \cdot c^i)(c^j \cdot c^j)}} \tag{eq.1}$$

Once **S** is constructed, we adopt a simple probabilistic method to generate an ensemble of 1,000 undirected networks. In detail, a link from risk $i$ to risk $j$ (and vice-versa) is introduced with a probability equal to their similarity i.e. increasingly alike risks are more likely to be linked. In addition, increasingly similar risks are expected to have stronger links i.e. the link weight is directly proportional to their similarity.

**Module identification**

Every module corresponds to a particular partition $N^* = \{n_1, ..., n_L\}$ of network $G(N, E)$. One way for identifying appropriate modules is to define a quality function $Q(G, N^*)$, where its value characterizes how good $N^*$ is a partition of $G$. Hence, the optimum set of modules can be obtained by maximizing $Q$.

To do so, we use an implementation of Blondel, Guillaume, Lambiotte, and Lefebvre (2008)'s algorithm, which utilizes a weighted variant of the Newman-Girvan modularity measure (Girvan & Newman, 2002; M. E. Newman, 2006), as an appropriate $Q$. This measure essentially accounts for the density of the links inside a given partition, compared to the links between the partitions. In the case of a weighted network, it is defined as (M. E. Newman, 2004):

$$Q = \frac{1}{2m} \sum_{i,j} \left( \mathbf{A}(i,j) - \frac{k_i k_j}{2m} \right) \delta(c_i, c_j) \tag{eq.2}$$

where $k_i$, defined as $\sum_{j, i \neq j} \mathbf{A}(i,j)$, reflects the sum of the weights of links attached to node $i$, $c_i$ corresponds to the module where node $i$ is assigned, Kronecker delta $\delta$ is 1 if $c_i = c_j$ and 0 otherwise, and finally, $m = \frac{1}{2} \sum_{i,j} \mathbf{A}(i,j)$. We note that there are cases where the particular null model deployed by this formulation (second term in summand of eq.2) is not suitable e.g. in the case of very broad degree distributions. Squartini and Garlaschelli (2011) provide a condition to assess the suitability of this null model, which states that if the maximum degree ($k_{\max}$) is lower than $\sqrt{2L}$, the null model in the original formulation can be used, where L is the total number of links in the network. In our case, $k_{\max} = 23.88$ and $\sqrt{2L} = 54.79$, satisfying the condition and in turn, confirming the suitability of this particular formulation.

Once the modules are obtained, we need to confirm that they contain meaningful information i.e. that their structure cannot be replicated by a random process. To do so, we use the methodology proposed by Clauset (2005) to generate random modular networks with the same number of modules. We then use the Normalized Mutual Information (NMI) measure (Danon et al., 2005) to compare the modules found in the risk network, with the ones found within its random counterpart. An NMI value of 0 indicates no similarity between the two networks, and a value of 1 if the modules are identical. By comparing an ensemble of 1,000 risk networks with their 1,000 artificial counterparts, we obtain an NMI value of 0.0749 (standard deviation is 0.0186), confirming the utility of the modules identified. Each module is visualized in Section 5, SI.

Lastly, we note that this particular formulation for $Q$ (eq.2) is subject to an intrinsic resolution limitation, which can bias the process of module identification (Arenas, Fernandez, & Gomez, 2008; Fortunato & Barthélemy, 2007; Nicolini, Bordier, & Bifone, 2017). The impact of this limitation can be severe as it can lead to the failure of identifying modules smaller than a given scale, resulting in modules which are composed of self-consistent, sub-modules. To evaluate whether a module is smaller than this scale, and thus subject to this limitation, Fortunato and Barthélemy (2007) used the number of links contained within a given module $s$, $l^s$, and $L$ to develop the following condition, $l^s < \sqrt{2L}$. Satisfying this condition means that module $s$ is composed of sub-modules, and is therefore not self-consistent. In our case, $l^1 = 307$, $l^2 = 255$, $l^3 = 114$, $l^4 = 143$ and $l^5 = 90$ – all are larger than $\sqrt{2L} = 54.79$. Hence, our results are robust against the resolution limitations of eq.2, and no sub-modules are contained within Modules 1-5.

**Evaluating Systemic Impact**

We use a simple 'Susceptible-Infected' model to evaluate the total number of risks triggered due to the manifestation of risk $i$. The state of each risk is defined as 'materialized' or 'non-materialized', recorded as $s = 1$ or $s = 0$ respectively. The algorithm for implementing the 'Susceptible-Infected' model is as follows: (1) select risk $i$ and switch its state from $s_i = 0$ to $s_i = 1$; (2) identify its neighboring risk(s) $j$ and (3) evaluate whether they are affected by the materialization of risk $i$. Step (3) is a probabilistic step, where a random value is drawn from a uniform distribution and is compared to the similarity between risk $i$ and $j$ – if the similarity is higher, the state of risk $j$ switches to 'materialized' i.e. $\mathbf{P}(s_j = 1 | s_i = 1) = \mathbf{A}(i, j)$. Once this procedure is completed – either because no more risks are left to be affected or because risk $i$ has no neighboring node(s) – the number of risks affected is summed and used to define the systemic impact of risk $i$. The process is then reiterated across all nodes. Results presented herein are an average from 1,000 independent runs.

The underlying assumption of this process is simple yet powerful: risks with increasingly similar characteristics are more likely to be triggered by similar cause(s). With that in mind, step (1) assumes that the conditions responsible for triggering risk $i$ have been met. Consequently, if risk $j$ is increasingly similar to risk $i$, the met conditions are also likely (but not guaranteed) to trigger risk $j$; the probability for doing so is determined in step (3). In this spirit, the converse argument is also true i.e. mitigating risk $i$ suggests that the conditions responsible for it have been treated, and therefore risk $j$ is less likely to occur, depending on the similarity between the two risks.

**From quantitative to qualitative classification of systemic impact**

The procedure to convert the quantitative results of systemic impact to the classification used in Figure 4 (i.e. 'High', 'Medium' and 'Low') is as follows: (1) evaluate the number of risks which have a reported 'High', 'Medium' and 'Low' independent impact, as found within the original data. This breaks down to 61, 58 and 24 risks respectively. For consistency, we preserve this decomposition, by (2) ranking risks in terms of their systemic impact and (3) assign the top 61 as 'High', the next 58 as 'Medium' and the remaining entries as 'Low', in terms of their systemic impact.

**Robustness of Results**

Our results heavily depend on the actual topology of the risk network, which in turn depends on the method used to determine risk similarity, in particular using Cosine Distance. Therefore, evaluating the dependency of our results to this particular similarity measure an important aspect, as one would hope for results to be robust against slightly different measures. To do so, we focus on two key outputs – (a) the evident mismatch between independent and systemic risk impact, and (b) the particular modular structure that characterizes the risk network – and how these may vary when different similarity measures are deployed for the generation of the risk network.

In general, similarity measures can be categorized in two classes (Lesot, Rifqi, & Benhadda, 2008): (i) Type 1, which considers only positive matches between existing attributes as contributors to the overall similarity between two vectors (i.e. a particular attribute is present in both vector A and B, hence they are increasingly similar); and (ii) Type 2, which considers both positive and negative matches, where the absence of a particular attribute further contributes to their similarity (i.e. a particular attribute is absent from both in vector A and B, hence they are increasingly similar). In this context, Type 2 measures are not suitable since negative matches do not necessary imply any similarity between two risks, due to the potentially infinite number of attributes that may be lacking in their respective characteristic vectors (Choi, Cha, & Tappert, 2010; Sneath & Sokal, 1973). Therefore, we will limit our robustness test to Type 1 similarity measures.

Type 1 similarity measures can be formalized using three key components: $a$, which refers to the number of features present in both vectors (i.e. positive matches); $b$, the number of attributes that exist in vector A and not in B, and $c$, the number of attributes that exist in vector B and not in A. Trivially, $b + c$ refers to the total number of attributes that exist in vector A (B) and absent from B (A). Given this formulation, we consider four widely used, Type 1 similarity measures (Choi et al., 2010):

$$SM_{Dice} = \frac{2a}{2a + b + c} \quad \text{(eq.3)}$$

$$SM_{Jaccard} = \frac{a}{a + b + c} \quad \text{(eq.4)}$$

$$SM_{Lance \& Williams} = 1 - \left(\frac{b + c}{2a + b + c}\right) \quad \text{(eq.5)}$$

$$SM_{Sorgenfrei} = \frac{a^2}{(a + b)(a + c)} \quad \text{(eq.6)}$$

With respect to finding (a) – mismatch between independent and systemic impact – we repeat the analysis described in Methods, Evaluating Systemic Impact. For every additional similarity measure tested, we generate the respective adjacency matrix and rerun the 'Susceptible-Infected' model for 1,000 independent runs. In general, the number of risks where Systemic Impact is greater, or equal to, their Independent Impact is consistent across all similarity measures, highlighting the robustness of finding (a) – see Table 4. Therefore, results related to finding (a) are robust.

**Table 4**: Number of risks that illustrate a particular mismatch between Systemic and Independent Impact

| Distance measure used to generate risk network | Number of risks whose Systemic Impact ≥ Independent Impact | Number of risks whose Systemic Impact < Independent Impact |
|---|---|---|
| Cosine | 96 | 47 |
| Dice | 95 | 48 |
| Jaccard | 95 | 48 |
| Lance & Williams | 95 | 48 |
| Sorgenfrei | 94 | 49 |

With respect to finding (b) – the existence of a particular modular structure – we repeat the analysis described in Methods, Module Identification. For every additional similarity measure tested, we first generate an ensemble of 1,000 networks. For each ensemble, we identify the most frequent module in which each risk is assigned to, and compare that with its respective module assignment obtained using Cosine Distance. Figure 5 maps the overall match between the cluster assignment obtained using Cosine Distance, and the additional similarity measures. In general, module assignment under Dice,

Jaccard, and Lance & Williams similarity measures is almost identical to the one obtained using Cosine Distance. This is not the case for Sorgenfrei, where the match is poor.

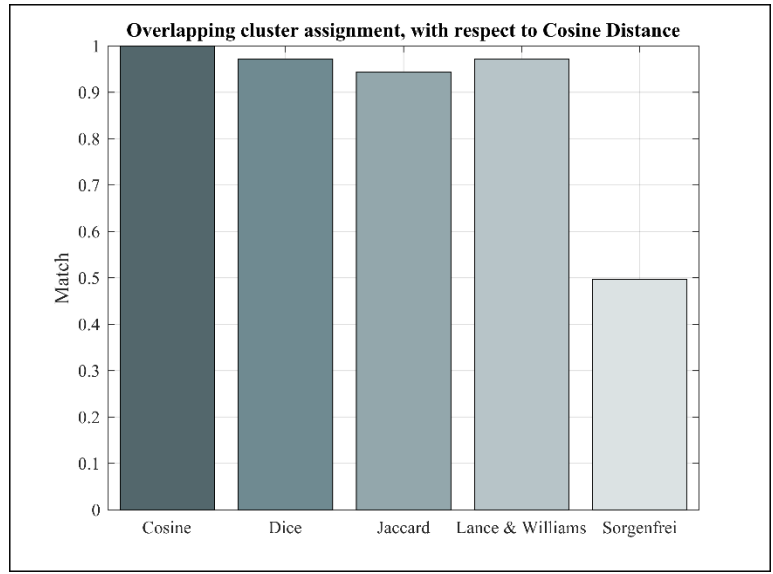

**Figure 5**: Match between module assignment obtained using Cosine Distance and other similarity measures (bar). A value of 1 indicates a complete match.

To identify the cause for this poor match, we performed a simple experiment to assess the sensitivity of each measure with respect to vector similarity. Consider vector A and B, the first being composed of 0s and the latter of 1s – at this point the similarity between A and B is 0. At each time step, vector A becomes incrementally similar to vector B by randomly choosing a 0 entry and switching its value to 1. Therefore, vector A becomes increasingly similar to vector B at every time step, until they become identical – at this point the similarity between A and B is 1. By monitoring the increase in similarity using different similarity measures, we can assess the sensitivity of each measure. In this case, superlinear behavior corresponds to heightened sensitivity, whilst sublinear behavior corresponds to reduced sensitivity – see Figure 6. Evidently, measures that have a good match in terms of reporting the same modules – Cosine, Dice, Jaccard; Lance & Williams; Figure 5, bar 1 to 3 – are ones that grow at least linearly with increased similarity, whilst measures that grow sublinearly – Sorgenfrei; Figure 5, bar 4 – perform poorly.

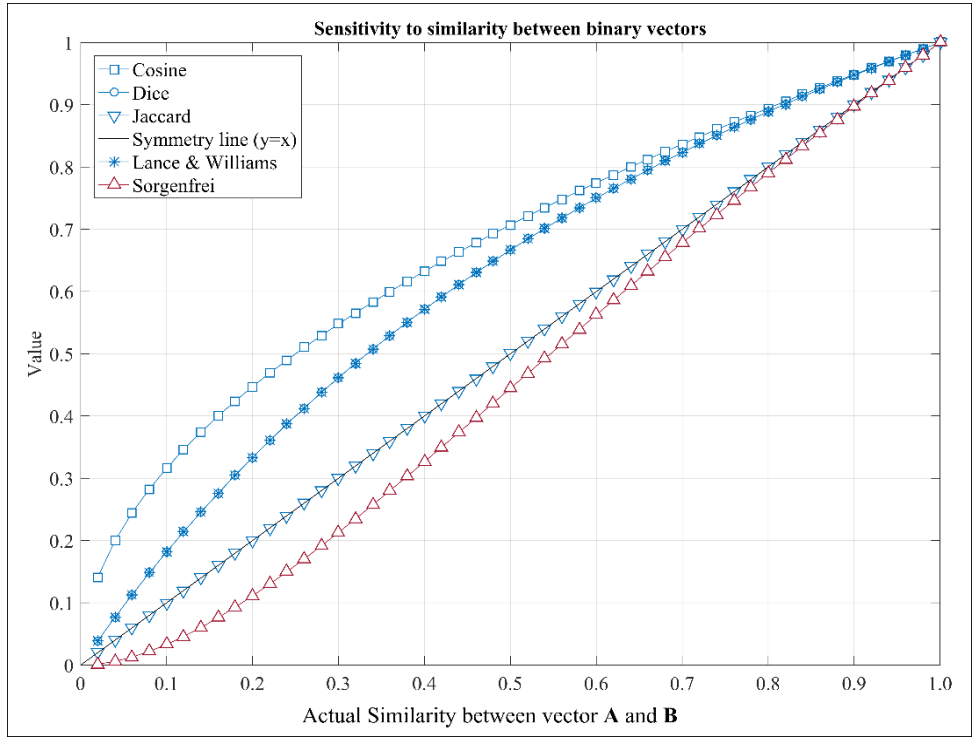

**Figure 6**: Sensitivity of each similarity measure (markers) as a function of increasingly similar vectors (x-axis). A symmetry line y-x is also included for reference.

To assess the generalizability of this statement, we define an additional measure designed to grow minimally with increased similarity $\left(SM_{test} = \frac{\min\left(\frac{a}{b+c}, a+b+c\right)}{a+b+c}\right)$ – see Figure 7a. As expected, the performance of this measure is exceedingly poor when considering the resulting cluster assignment, in relation to the ones obtained using Cosine Distance (Figure 7b).

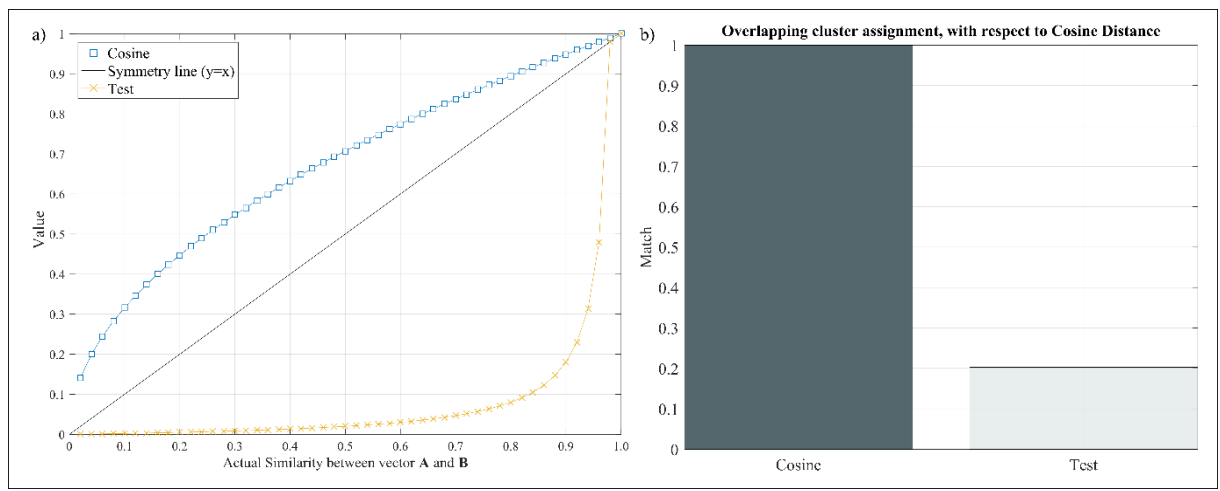

**Figure 7**: subplot (a) and (b) are the same as Figure 6 and Figure 5, with an explicit focus on comparing the newly introduced similarity measure (x marker) with Cosine Distance (square markers),

In conclusion, this section tests the dependency of the reported results with respect to the adopted similarity measure. The focus is on: (a) the evident mismatch between independent and systemic risk impact, and (b) the particular modular structure that characterizes the risk network. Both (a) and (b) are robust against the use of similar similarity measures, as shown in Table 4 and Figure 5 respectively. However, the robustness of (b) has an additional caveat – the measure used to evaluate similarity grow at least linearly with respect to the number of shared characteristics (Figure 6). Considering the nature of the data examined herein, this is a reasonable expectation as every additional positive match between two characteristic vectors contributes to the similarity of their respective risks.

## Additional information

### Availability of data and materials
Additional data supporting the conclusions of this article is included within the article and its additional files.

### Declaration of Interest
CE and NA were partly, and CC fully, employed by ORIC International, a non-profit organization in the (re)insurance and asset management sector. The authors alone are responsible for the content and writing of the paper.

### Author Contribution
CE and NA designed the experiment; CC provided the dataset; CE developed the model, analyzed the data, prepared figures and wrote the manuscript; CE, NA and CC reviewed and approved the manuscript.

### Funding
This work was partly funded by ORIC International (CE, NA and CC) and an EPSRC Doctoral Prize Fellowship (CE).

### Acknowledgements
We are grateful to Jenna Anders of ORIC International for useful comments, discussions and support.